\documentclass[]{emulateapj}

\usepackage{epsfig, natbib, graphicx, color}

\newcommand{\rhogas}{\rho_{gas}}
\newcommand{\keV}{\mbox{keV}}
\newcommand{\Rf}{R_{500}}
\newcommand{\Tmw}{T_{mw}}
\newcommand{\Mgas}{M_{gas}}

\newcommand{\Mpc}{\mbox{Mpc}}

\newcommand{\msun}{M_\odot}
\newcommand{\bmyayo}[1]{\mathbf{#1}}
\newcommand{\bx}{\bmyayo{x}}

\newcommand{\nn}{\nonumber}

\newcommand{\avg}[1]{\left\langle #1 \right\rangle}

\newcommand{\yx}{y_x}
\newcommand{\ysz}{y_{sz}}
\newcommand{\ssz}{\sigma_{sz}}
\newcommand{\sx}{\sigma_x}
\newcommand{\sszx}{\sigma_{sz|x}}

\newcommand{\beq}{\begin{equation}}
\newcommand{\eeq}{\end{equation}}
\newcommand{\beqa}{\begin{eqnarray}}
\newcommand{\eeqa}{\end{eqnarray}}

\newcommand{\be}{\begin{equation}}
\newcommand{\ee}{\end{equation}}

\newcommand{\bea}{\begin{eqnarray}}
\newcommand{\eea}{\end{eqnarray}}

\newcommand{\Ysz}{Y_{SZ}}

\newcommand{\lkhd}{{\cal L}}

\newcommand{\planck}{\emph{Planck}}
\newcommand{\chandra}{\emph{Chandra}}
\newcommand{\xmm}{\emph{XMM}}
\newcommand{\rosat}{\emph{ROSAT}}

\def\***#1{\textsf{\textbf{#1}}}

\shortauthors{ROZO et al.}
\shorttitle{$\Ysz$--$Y_X$ from {\it Planck} and {\it Chandra}}

\begin{document}
\title{The $\Ysz$--$Y_X$ Scaling Relation as Determined from {\it Planck} and {\it Chandra}}
\author{Eduardo Rozo\altaffilmark{1,2}, Alexey Vikhlinin\altaffilmark{3,4}, Surhud More\altaffilmark{1}}
\altaffiltext{1}{Kavli Institute for Cosmological Physics, Chicago, IL
 60637, USA}
 \altaffiltext{2}{Department of Astronomy, University of Chicago, Chicago, IL
 60637, USA}
 \altaffiltext{3}{Space Research Institute (IKI), Profsoyuznaya 84/32, Moscow 117810, Russia} 
 \altaffiltext{4}{Harvard-Smithsonian Center for Astrophysics, 60 Garden St., Cambridge, MA 02138, USA}
 
\begin{abstract}
  SZ clusters surveys like \planck, the South Pole Telescope, and the
  Atacama Cosmology Telescope, will soon be publishing several hundred
  SZ-selected systems.  The key ingredient required to transport the
  mass calibration from current X-ray selected cluster samples to
  these SZ systems is the $\Ysz$--$Y_X$ scaling relation.   We
  constrain the amplitude, slope, and scatter of the $\Ysz$--$Y_X$
  scaling relation using SZ data from \planck\ and X-ray data from
  \chandra.  We find a best fit amplitude of $\ln (D_A^2\Ysz/CY_X) =
  -0.202 \pm 0.024$ at the pivot point $CY_X=8\times 10^{-5}\ \Mpc^2$.
  This corresponds to a $\Ysz/Y_X$-ratio of $0.82\pm
  0.024$, in good agreement with X-ray expectations after including
  the effects of gas clumping. The slope of the relation is
  $\alpha=0.916\pm 0.032$, consistent with unity at $\approx
  2.3\sigma$.  We are unable to detect intrinsic scatter, and find no
  evidence that the scaling relation depends on cluster
  dynamical state.
\end{abstract}
\keywords{clusters
}

\section{Introduction}

The Sunyaev--Zeldovich effect --- i.e.\ the prediction that the hot
intra-cluster gas of massive galaxy clusters will scatter CMB photons
to produce a unique frequency signature --- traces back to 1972
\citep{sunyaevzeldovich72}.  Despite early detections of this effect
\citep{gullnorthover76}, it is only recently that instrumentation
advances have made large scale SZ searches feasible.  There are three
such ongoing surveys: one carried out with the South Pole Telescope
\citep[SPT][]{carlstrometal11}, one with the Atacama Cosmology
Telescope \citep[ACT][]{fowleretal07}, and one using the \planck\
satellite \citep{planck11_mission}.  All three surveys have published
initial cluster samples
\citep{vanderlindeetal10,marriageetal11,planck11_earlysample}, and are
expected to eventually publish hundreds of SZ selected systems out to
$z=1$ and beyond.  If the scaling relation between the integrated SZ
signal and cluster mass can be adequately calibrated, then these
cluster samples may be used to constrain the growth of large scale
structure in the universe, which can in turn provide critical
constraints on dark energy models \citep[e.g.][]{holderetal01}.
Indeed, both SPT and ACT have already derived cosmological constraints
from their current samples
\citep{vanderlindeetal10,sehgaletal11,bensonetal11}.

Unfortunately, at this time no
single method provides masses of individual clusters with a required
$\sim 5\%$ uncertainty. X-ray hydrostatic measurements are
observationally expensive and limited to a subset of relaxed systems.
Numerical simulations suggest that hydrostatic mass estimates can 
be biased low anywhere between 10\% 
to 30\% \citep{nagaietal07a,piffarettivaldarnini08,jeltemaetal08,lauetal09,meneghettietal10,rasiaetal12}.
Weak lensing estimates are in principle unbiased, but specific implementations can result in
$\approx 5\%-10\%$ biases depending on the fitting 
procedure \citep[typically biased high when using NFW fitting formulae][]{beckerkravtsov10,meneghettietal10,rasiaetal12},
and exhibit large intrinsic scatter.
In addition, without the help from X-ray data, the derived SZ signal from the current generation surveys has a comparatively large scatter at a fixed 
mass \citep{bensonetal11,planck11_earlysample}.  Mass calibration using velocity dispersions is also 
possible \citep[e.g.][]{evrardetal08,sifonetal12}, but subject to velocity bias.
Therefore, the precise calibration of the cluster masses from SZ surveys will be possible 
only through a combined analysis of the SZ, X-ray, and optical data. Consequently, all three of the on-going SZ experiments 
are devoting significant effort for extensive X-ray and weak lensing follow-up of the SZ-selected 
systems \citep{anderssonetal10,foleyetal11,planck11_followup,planck11_validation,menanteauetal11}.

Another possibility, however, is simply to transfer the existing calibration of the
mass-proxy relations in the X-ray samples \citep[e.g.][]{arnaudetal07,allenetal08,vikhlininetal09,arnaudetal10,mantzetal10b}
using the correlation of the
cluster SZ parameters with the X-ray proxies \citep[e.g.][]{bensonetal11}.
An important step in this process is to verify a tight
correlation between the integrated SZ signal, $\Ysz$, and its X-ray
analog, $Y_X=T_{X}\times M_{{\rm gas},X}$ \citep[][see section \ref{sec:ratio_def} for
more precise definitions of $\Ysz$ and $Y_X$]{kravtsovetal06}. In
addition to transferring the mass calibration, the $\Ysz-Y_{X}$
relation can be used to verify some key predictions of the structure
formation theory which is used for applying the cosmological models to
the cluster data. First, numerical simulations consistently show that
both $Y_{SZ}$ and $Y_{X}$ have low scatter at a fixed mass
\citep[$\approx 10\%-15\%$,][]{nagai06,kravtsovetal06,staneketal10,fabjanetal11,kayetal11}.
Thus, we can expect the $\Ysz$--$Y_X$
relation to have very low scatter, especially since these parameters
approximate the same physical quantity, the thermal energy in the
cluster gas. Second, these same simulations predict nearly self-similar
evolution in the $Y-M$ relation for both $Y_{X}$
and $\Ysz$, so we
expect the $\Ysz$--$Y_X$ relation to have nearly the same normalization
at both low and high redshifts.

The $\Ysz$--$Y_{X}$ relation has been reported in several recent works
--- by \cite{anderssonetal10} at high-$z$ using SPT and
\emph{Chandra}, and at low-$z$ by \cite{planck11_local}
using Planck and \emph{XMM-Newton}. The results are generally in line
with the theoretical expectations --- $\Ysz$ follows $Y_{X}$ with a
small overall offset. However, there are several issues which warrant
further study. The overall offset between $Y_{X}$ and $\Ysz$ is
expected because real clusters are non-isothermal and slightly
non-uniform.  Given the overall $\rho^2$ weighting of X-ray
emission versus the $\rho$ weighting of the SZ-signal,
one naturally expects the two quantities to differ (see below
for further details).   
Based on the analysis of \emph{XMM-Newton}
temperature profiles in a ``representative'' sample of clusters,
\citet{arnaudetal10} predicted $\Ysz/Y_{X}=0.924$. The mean offset in
the SPT/\emph{Chandra} sample of \cite{anderssonetal10} is somewhat
below this prediction, $\Ysz/Y_{X}=0.82\pm 0.07$. The mean offset in
the Planck/\emph{XMM-Newton} sample is $1\,\sigma$ above,
$\Ysz/Y_{X}=0.95\pm0.03$ \citep{planck11_local}.  While the difference
in the $\Ysz/Y_X$ ratio is relatively large, it is not significant because
of the large errors in the \citet{anderssonetal10} measurement,
so additional measurements that reduce the statistical uncertainty
of \chandra\ estimates of the $\Ysz/Y_X$ ratio are of great 
interest.   This is particularly true in light of known differences
in instrument calibration between \chandra\ and \xmm\, so having
a low-redshift calibration of this ratio using \chandra\ is of significant interest.
Finally, the scatter in the \planck/\xmm\ $\Ysz$--$Y_X$
  relation is $23\% \pm 2\%$, which is significantly larger than the 
  expected $\approx 10\%$ scatter, so confirmation of this observation
  is important.
  
The goal of this paper is to measure the local $\Ysz-Y_{X}$ relation
independently using overlapping clusters in the Planck and
\citet{vikhlininetal09} \emph{Chandra} sample. This is the first step
towards \emph{Chandra}-based calibration of the local $\Ysz-M$
relation. We defer the actual calibration of the $\Ysz$--$M$ relation
to an upcoming paper, where we simultaneously consider several the
$\Ysz$--$M$ scaling relation as calibrated by several groups
\citep{rozoetal11c}.  In this context, we emphasize that the Chandra
X-ray results quoted here are derived from the gas temperature and gas
density models for individual clusters taken from
\citet{vikhlininetal09}.  Consequently, this work and that of
\citet{vikhlininetal09} are fully self-consistent in all definitions.

Throughout this work, we assume a flat $\Lambda$CDM cosmology with
$\Omega_m=0.3$ and $h=0.72$.  This convention matches that of \citet{vikhlininetal09}.

%%%%%%%%%%%%%%%%%%%%%%%%%%%%%%%%%%%%%%%%
%%%%%%%%%%%%%%%%%%%%%%%%%%%%%%%%%%%%%%%%
%%%%%%%%%%%%%%%%%%%%%%%%%%%%%%%%%%%%%%%%
%%%%%%%%%%%%%%%%%%%%%%%%%%%%%%%%%%%%%%%%
%%%%%%%%%%%%%%%%%%%%%%%%%%%%%%%%%%%%%%%%
%%%%%%%%%%%%%%%%%%%%%%%%%%%%%%%%%%%%%%%%

\subsection{What We Mean by the $\Ysz/Y_X$ Ratio}
\label{sec:ratio_def}

The $\Ysz$ signal of a galaxy cluster is related to the total pressure
of the intra-cluster gas within the cluster's volume via
\be
D_A^2 \Ysz = C \int dV \rhogas(\bx) T(\bx)
\label{eq:szint}
\ee
where $D_A$ is the angular diameter distance,  and $C$ is a constant that switches the units back and forth from $\msun \keV$
to $\Mpc^2$.
The constant $C$ is given by
\be
C=\frac{\sigma_T}{m_e c^2}\frac{1}{\rho_{\rm gas}/n_e} = 1.406\times \frac{ 10 ^{-5}\ \Mpc^2 }{10^{14}\ \keV\msun},
\ee
where $\sigma_T$ is the Thompson cross-section, and $\rho_{\rm
  gas}/n_e=1.149\,m_p$ for fully ionized plasma with the cosmic He
abundance and abundance of heavier elements set at
solar. \citet{arnaudetal10} assume a heavy-element abundance that is $0.3$ solar,
which differs from the above value by less than $1\%$, a difference that is completely
negligible for our purposes.

From equation \ref{eq:szint}, it is apparent that the combination
$D_A^2\Ysz$ will scale with $CY_X$ where $Y_X = \Mgas \times T_X$ (see
section \ref{sec:xray_data} for our precise definition of $\Mgas$ and
$T_X$).  Thus, we expect that the $\Ysz/Y_X$ ratio is a dimensionless
number that is close to unity.  Throughout this work, we will often
speak about the $\Ysz/Y_X$ ratio in this fashion.  In practice,
however, $\Ysz$ and $Y_X$ are cluster observables that are related via
a probability distribution $P(\Ysz|Y_X)$.  Here, we assume
$P(\Ysz|Y_X)$ is a log-normal distribution, which takes the form
\be
\avg{\ln \left(D_A^2\Ysz\right) |Y_X} = a+\alpha \ln CY_X.
\ee
When we speak of ``the $\Ysz/Y_X$ ratio'', we mean the quantity
$\avg{\ln \left(D_A^2\Ysz\right) |Y_X} - \ln CY_X$ evaluated at
the pivot point of the sample.  Note that if we select units such that
$CY_X=1$ at the pivot point, then the $\Ysz/Y_X$ ratio is just the
amplitude $a$.  
We will work primarily in units of $CY_X = 8\times 10^{-5}\ \Mpc^2$,
corresponding to the pivot point for the \chandra--\planck\ cluster sample 
(see section \ref{sec:data}).  
For brevity, we will often simply write $\ln(\Ysz/Y_X)$ ---
as opposed to the full $\avg{\ln D_A^2\Ysz |Y_X} - \ln CY_X$
expression --- to denote the $\Ysz/Y_X$ ratio.

%%%%%%%%%%%%%%%%%%%%%%%%%%%%%%%%%%%%%%%%%%%%%%%%
%%%%%%%%%%%%%%%%%%%%%%%%%%%%%%%%%%%%%%%%%%%%%%%%

\section{X-ray Predictions}
\label{sec:predictions}

Equation \ref{eq:szint} allows one to compute the $\Ysz$ signal of a
galaxy cluster based on the gas and temperature profiles of galaxy clusters.  With
sufficiently high quality X-ray data, both of these profiles can be
observationally constrained, allowing one to directly predict the
$\Ysz$ signal of a galaxy cluster from the X-ray data.  Using this
method, \citet{arnaudetal10} predicted that
$D_A^2\Ysz/CY_X=0.924$ or $\ln(\Ysz/Y_X) = - 0.08$ when $T_X$ is
defined within a $[0.15,0.75]\Rf$ aperture.
Using the \citet{vikhlininetal06} \emph{Chandra} results for local
relaxed clusters, we predict nearly the same ratio. Specifically, the
mass-weighted temperature
\be
\Tmw = \frac{\int dV\ \rhogas(\bx)T(\bx)}{\int dV\ \rhogas(\bx)},
\label{eq:Tmw}
\ee
is related to the spectroscopic temperature $T_X$ via
$\Tmw/T_X=0.928$, where $T_{X}$ is measured in the $[0.15-1]\,R_{500}$
aperture. Using the definition for $\Tmw$, we see that we can rewrite
equation \ref{eq:szint} via
\be
D_A^2 \Ysz = C \Mgas \Tmw,
\ee
from which we find $\ln(\Ysz/Y_X) = - 0.07$.

It is potentially a concern that the \citet{arnaudetal10} work is
based on a representative cluster sample of X-ray luminous clusters while
\citet{vikhlininetal06} include only relaxed, cool core
clusters. However, empirically, there is very little difference in
$\Ysz/Y_{X}$ for relaxed and unrelaxed clusters judging either from
the X-ray predictions \citep[Fig.9 in][]{arnaudetal10} or the actual
measurements \citep[see Fig.4 in][and our results
below]{planck11_local}. Another point of concern is slightly different
apertures for $T_{X}$ used in the \emph{XMM} and \emph{Chandra}
samples. One expects that mean (emission-weighted) temperatures in the
$[0.15-1]\,R_{500}$ and $[0.15-0.75]\,R_{500}$ annuli should be close
because only a small fraction of the total X-ray flux comes from
outside $0.75\,R_{500}$. 
Indeed, the $T_X([0.15,0.75]R500)/T_X([0.15,1]R500)$ ratios in the Chandra
sample of \citet{vikhlininetal09} are within $\pm 0.015$ of a
mean value of 0.985.   An analysis of the deeper pointings among the ESZ-XMM sample 
(G.Pratt, private comm.) shows a similar value, even though the temperature ratio for the 
REXCESS sample \citep{prattetal09} was somewhat lower on average ($\approx 0.95$), and showed
a larger spread $0.85-1$.  
We proceed accordingly, and ignore the difference in the definition of TX between the two works. 
With this caveat
in mind, we conclude that both \emph{XMM} and \emph{Chandra} predict
$\ln(\Ysz/Y_X) \simeq -0.075$ because of non-isothermal cluster
temperature profiles.

There is, however, an additional complication that modifies these
expectations. Specifically, when one estimates $\Mgas$ from X-ray
data, one uses the fact that the surface brightness $S\propto
\rhogas^2$ to estimate $\avg{\rhogas^2}$, averaged in radial annuli.
For a uniform, spherically symmetric distribution
$\avg{\rhogas^2}=\avg{\rhogas}^2$, so one can recover the gas density
profile.  In practice, however, one will typically have
$\avg{\rhogas^2} > \avg{\rhogas}^2$.  The clumping factor $Q^2$ is
defined via
\be Q^2= \frac{\avg{\rhogas^2}}{\avg{\rhogas}^2}, 
\ee
so that $\Mgas^{obs}=Q\Mgas^{true}$.  By definition,
$Y_X=\Mgas^{obs}T_X$, so the $\Ysz$ to $Y_X$ ratio takes the form
\be
\frac{D_A^2 \Ysz}{CY_X} = \frac{1}{Q}\frac{\Tmw}{T_X}.
\label{eq:Tratio}
\ee
In practice, the clumping factor $Q$ depends on radius, and the factor
$1/Q$ characterizes the mean clumping correction over the radius
$\Rf$.  Based on this discussion, we modify our X-ray expectation by a
factor of $Q^{-1}$.  If there is no substructure masking in the X-ray
analysis, $Q$ can be very large
\citep[e.g. $Q=1.16$,][]{mathiesenetal99}, though analysis of
numerical simulations that include substructure masking achieve
significantly lower values of $Q$. Based on the analysis of
\citet{nagaietal07a}, we set $Q=1.05\pm 0.05$, where the error is
set to the size of the correction to account for systematic uncertainties in the
simulation physics.  The corresponding expectation for the 
$\Ysz/Y_X$ ratio is $\ln (D_A^2\Ysz/CY_X) = -0.125\pm 0.05$.

Finally, we note that $T_{X}$ measured in the X-ray can be subject to
a systematic uncertainty, related to, e.g., absolute calibration of
the X-ray telescope effective area. Assuming for simplicity that this
results in a uniform bias of observed temperatures,
$T_{X}^{\rm(meas)}=A\,T_{X}^{\rm(true)}$, the measured ratio of the
$Y$'s can be factorized as
\be
\frac{D_A^2 \Ysz}{CY_X} = \frac{1}{A\,Q}\frac{\Tmw}{T_X}.
\label{eq:Tratio2}
\ee
Discussion of the calibration-related uncertainties in $T_{X}$ is
beyond the scope of this paper and we proceed assuming $A=1$. However,
it is important to keep in mind a possibility of a few percent
systematic error in the X-ray temperature measurements with the
current calibration of the \emph{XMM-Newton} and \emph{Chandra}
instruments. To give a handle of possible deviations of $A$ from
unity, we note that within the sample we consider in this work, the
average ratio of \emph{XMM} and \emph{Chandra}-measured temperatures
for the same cluster is $\ln (T_{XMM}/T_{Chandra} ) = -0.08$ 
(for a detailed comparison of \xmm\ and \chandra\ data on a cluster-by-cluster
basis see Rozo et al. 2012, in preparation).\footnote{The quoted ratio does {\it not} correct
for the difference between the temperature definitions used in the relevant
\xmm\ and \chandra\ works as noted earlier.}

%%%%%%%%%%%%%%%%%%%%%%%%%%%%%%%%%%%%%%%%
%%%%%%%%%%%%%%%%%%%%%%%%%%%%%%%%%%%%%%%%
%%%%%%%%%%%%%%%%%%%%%%%%%%%%%%%%%%%%%%%%
%%%%%%%%%%%%%%%%%%%%%%%%%%%%%%%%%%%%%%%%
%%%%%%%%%%%%%%%%%%%%%%%%%%%%%%%%%%%%%%%%
%%%%%%%%%%%%%%%%%%%%%%%%%%%%%%%%%%%%%%%%

\section{Data}
\label{sec:data}

\subsection{Cluster Sample}

%%%%%%%%%%%%%%%%%%%%%%%%%%%%%%%%%%%%%%%%%%%%%%%%
%%%%%%%%%%%%%%%%%%%%%%%%%%%%%%%%%%%%%%%%%%%%%%%%

\begin{figure} 
\hspace{-0.2in} \includegraphics[width=3.6in]{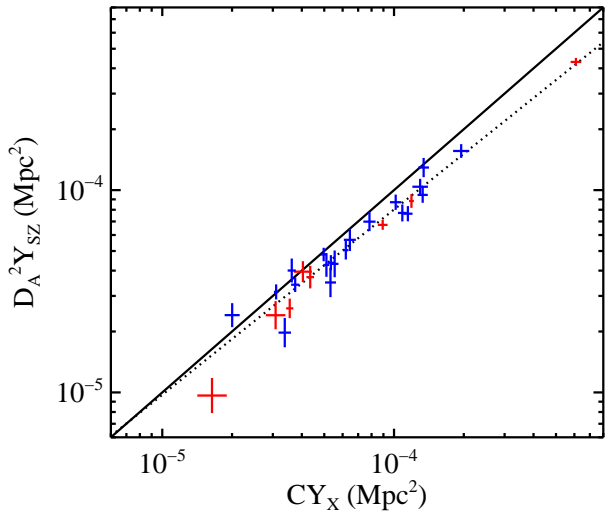}
\caption{The relation between the Planck $D_A^2Y_{SZ}$ and
  \emph{Chandra} $CY_X$ measured in the same aperture, $R_{500}$
  provided by \emph{XMM}-Newton mass estimates.  Blue and 
  red crosses  represent dynamical relaxed and merging clusters,
  respectively (per the \citet{vikhlininetal09} classification). The solid line
  shows a 1-1 relation, while the dotted line is our best fit relation.
  The amplitude at the pivot point corresponds to 
   $D_A^2Y_{SZ}/CY_{X}=0.82$.
   }
\label{fig:yx-ysz}
\end{figure}

The cluster sample employed in this work is a subset of the Early SZ
(ESZ) sample published in \citet{planck11_earlysample}.  The ESZ
sample is comprised of 189 cluster candidates identified in a blind,
multi-frequency search of the all-sky maps obtained from the first ten
months of observations of the \planck\ mission
\citep{planck11_mission}.  The threshold for cluster detection was set
at a signal-to-noise $(S/N) \geq 6$.  This ESZ sample was searched for
archival $\xmm$ data, and a sub-sample of 88 galaxy clusters with
\xmm\ data were identified.  Of these, 62 clusters were processed in
time for publication, as detailed in \citet{planck11_local}.  Starting
from this sub-sample of 62 galaxy clusters, we identified 28 clusters
with \chandra\ data that had already been analyzed using the data
reduction pipeline of \citet{vikhlininetal09}.  The analysis presented
here is based on this sub-sample of $28$ galaxy clusters, which is
summarized in Table \ref{tab:clusters}.  
20 of these are relaxed galaxy clusters, and 8 are merging systems.
A plot of the $\Ysz$ vs $Y_X$
data for the galaxy clusters is shown in Figure
\ref{fig:yx-ysz}.

\subsection{Planck Data}

For each of our galaxy clusters, \citep{planck11_local} estimated the
integrated Compton parameter $\Ysz$ using a matched-filter algorithm.
The matched filter assumes an \citet{arnaudetal10} pressure
profile, which is used to predict the CMB temperature perturbation
$\delta T(\theta)$ of each galaxy cluster in each of the \planck\
frequency bands.  In order to perform the fit, the clusters are first
assigned a putative $\Rf$ radius based on the \xmm\ X-ray data and the
$M$--$Y_X$ scaling relation of \citet{arnaudetal10}.  The amplitude of
the matched-filter, which can be expressed in terms of the integrated
Compton parameter $\Ysz$, is then fit using the temperature data
within a projected aperture of $5\Rf$.  We note that the shape of the
matched filter is held fixed in the \citet{planck11_local} analysis,
but their systematics tests demonstrate that varying the shape of the
filter within tolerable levels does not significantly impact the
recovered scaling relations.  The median error in the recovered $\Ysz$
values for our cluster sample is $11\%$, and systematic uncertainties
due to instrumental beam size and radio point source contamination are
all expected be at the few percent level at most.  Note that because
\citet{planck11_local} assumed $h=0.7$ while we assume $h=0.72$, we
must rescale the $D_A^2\Ysz$ values in \citet{planck11_local}
appropriately.  Further, the Planck team uses the redshifts listed for
the clusters in the NED and SIMBAD databases while we use the
redshifts of the dominant central galaxies. We believe the latter
approach is more robust because it allows one to avoid contamination
by projection effects in some cases. Therefore, we rescaled all
angular diameter distances to our adopted $z$ values. The cluster
redshifts, along with the rescaled $\Ysz$ values, are detailed in
Table \ref{tab:clusters}.

%%%%%%%%%%%%%%%%%%%%%%%%%%%%%%%%%%%%%%%%%%%%%%%%
%%%%%%%%%%%%%%%%%%%%%%%%%%%%%%%%%%%%%%%%%%%%%%%%

%%%%%%%%%%%%%%%%%%%%%%%%%%%%%%%%%%%%%%%%%%%%%%%%
%%%%%%%%%%%%%%%%%%%%%%%%%%%%%%%%%%%%%%%%%%%%%%%%

\subsection{Chandra Data}
\label{sec:xray_data}

Our analysis is based on the sub-sample of ESZ galaxy clusters from
\citet{planck11_earlysample} that have \chandra\ data that was reduced
as part of the work in \citet{vikhlininetal09}, described in full
detail in \citet{vikhlininetal05,vikhlininetal06}.  Here, we provide
only a brief summary.  We note that not all galaxy clusters in our
analysis are included in the \citet{vikhlininetal09} sample. These are
the clusters which were analyzed by \citet{vikhlininetal09} but not
included in the flux-limited sample used in their cosmological
analysis. The full list of clusters employed in our analysis is presented
in Table \ref{tab:clusters}.

We use the available \chandra\ data to estimate $Y_X=\Mgas\times T_X$
for each of the clusters in our sample, where $\Mgas$ is the total gas
mass within the radius $\Rf$ employed by \citet{planck11_local}, 
and $T_X$ is the spectroscopic temperature within an annulus $R\in[015,1]\Rf$.  
To estimate $\Mgas$, we rely on analytic fits to the gas density profiles
of individual clusters. 
The fits were based on
the X-ray brightness profiles in the 0.7--2 keV energy band measured
with \emph{Chandra} and \emph{ROSAT} PSPC (see \citet{vikhlininetal09}
for details). All detectable X-ray substructures were masked out from
the profile measurements to minimize the impact of gas clumping. The
X-ray flux in the $0.7-2\ \keV$ band is very insensitive to the plasma
temperature so long as $T\gtrsim 2\ \keV$, as in the case of the
sample under consideration.  Nevertheless, we apply the appropriate
temperature and metallicity corrections to the X-ray profile to
estimate $\Mgas$.  The mean cluster temperature
reported by \citet{vikhlininetal09} corresponds to the
single-temperature fit to the X-ray spectrum in the $0.6-10\ \keV$
band extracted within an annulus $[0.15,1]\Rf$. The median error in
the recovered $Y_X$ values for our cluster sample is $3\%$, nearly
identical to the median error in the corresponding \xmm\ X-ray data in
\citet{planck11_local}.

\subsection{On the Importance of Apertures}
\label{sec:aperture}

As already mentioned in section \ref{sec:ratio_def}, we ignore any
radial dependence on the $\Ysz$--$Y_X$ scaling relation.  In other
words, we are explicitly assuming that the parameters $a$ and $\alpha$
are insensitive to the radius at which both $\Ysz$ and $Y_X$ are
evaluated.  This is clearly not true over arbitrarily large radial
ranges, but we expect any evolution over the range of radii probed to
be at the $\lesssim 1\%$ level, so long as both $\Ysz$ and $Y_X$ are
always measured within the same aperture.  Since we rely on the SZ
data from \citet{planck11_local}, we have to use the same metric
apertures adopted in that work.  Specifically, to estimate $Y_X$, we
first measure $\Mgas$ within the assigned $\Rf$ aperture, and multiply
this by the estimated temperature $T_X$ as quoted in
\citet{vikhlininetal09}.   We do not remeasure $T_X$ since the X-ray
temperature is robust to modest changes in the assigned $\Rf$
radius of a cluster.
A relation between \planck-measured $Y_{SZ}$
and the \emph{Chandra} $Y_{X}$ values is shown in
Fig.~\ref{fig:yx-ysz}.   There is an obvious, tight correlation 
that is nearly linear, with no obvious offset between merging and
relaxed systems.

%%%%%%%%%%%%%%%%%%%%%%%%%%%%%%%%%%%%%%%%%%%%%%%%%%%
%%%%%%%%%%%%%%%%%%%%%%%%%%%%%%%%%%%%%%%%%%%%%%%%%%%
%%%%%%%%%%%%%%%%%%%%%%%%%%%%%%%%%%%%%%%%%%%%%%%%%%%
%%%%%%%%%%%%%%%%%%%%%%%%%%%%%%%%%%%%%%%%%%%%%%%%%%%

\section{Likelihood Model}

Because we will be assuming log-normal models throughout, we begin
by defining $\yx$ and $\ysz$ via
\bea 
\yx & = & \ln \left( \frac{C Y_X}{8\times 10^{-5}\ \Mpc^2} \right) \\
\ysz & = & \ln \left( \frac{ D_A^2\Ysz}{8\times 10^{-5}\ \Mpc^2} \right).
\eea
We divide both $Y_X$ and $\Ysz$ by $8\times 10^{-5}\ \Mpc^2$ so that both quantities are still measured
in the same units, while simultaneously shifting the pivot point of the $\ysz$--$\yx$ scaling
relation to ensure the amplitude and slope of the best-fit $\Ysz$--$Y_X$ relation are nearly uncorrelated.

For each cluster, we assume that the observed $\yx$ value is given by
\be
\yx = \bar \yx + \delta_y
\label{eq:model}
\ee
where $\bar \yx$ is the true value, and $\delta_y$ is a random gaussian fluctuation of zero mean
and $\avg{\delta_y^2}=\sx^2$, where $\sx$ is the assigned observational error.  A similar expression
holds for $\ysz$ in terms of the true $SZ$ signal $\bar \ysz$.    The true values $\bar \ysz$ and $\bar \yx$
are themselves related via a probability distribution $P(\bar\ysz|\bar \yx)$ that characterizes the
scaling relation.  We assume this is a log-normal distribution of mean
\be
\avg{\ysz|\yx} = a + \alpha \yx
\label{eq:powerlaw}
\ee
and variance $\sszx^2$, so that $\sszx$ is the intrinsic scatter in the $\Ysz$--$Y_X$ relation.
We assume this scaling relation is redshift independent, so $a$, $\alpha$, and $\sszx$ are all
redshift independent.
Note that because we have chosen our units to ensure that $\yx=0$ at $CY_X=8\times 10^{-5}\ \Mpc^2$,
the amplitude $a$ is the $\Ysz/Y_X$ ratio at this scale.

The probability that a galaxy cluster has observed values $\yx$ and $\ysz$
can be written as
\bea
\hspace{-0.25in} P(\ysz,\yx) & = & \int d\bar \yx d\bar \ysz\ P(\ysz,\yx|\bar \ysz,\bar \yx) P(\bar \ysz,\bar \yx) \\
	& \hspace{-0.8in} = &  \hspace{-0.4in}  \int d\bar \yx d\bar \ysz\ P(\ysz|\bar \ysz)P(\yx|\bar \yx) P(\bar \ysz|\bar \yx) P(\bar \yx), \label{eq:p1}
\eea
where we are using the fact that the observational errors in $\yx$ and $\ysz$ are uncorrelated.  In principle, there can be a small amount
of covariance induced by aperture effects.  We discuss this effect in more detail below, but the main conclusion is that the effect is negligible
for our purposes, and so we will ignore this effect in our discussion.

Note the probability $P(\ysz,\yx)$ depends on the distribution $P(\bar \yx)$ of the cluster sample under consideration,
which depends on cluster selection effects.  We consider two possibilities, namely a flat prior for
the value $\bar \yx$, and a power-law distribution in $Y_X$, corresponding to $P(\bar \yx) = \exp(-\beta \yx)$.
The slope $\beta$ can then be set to the expected slope for a WMAP7 cosmology.
In practice, neither is exact, but it can help give us a sense of the level of systematic uncertainty in our
data due to selection effects.

Setting $P(\bar \yx)$ to a constant, integration of equation \ref{eq:p1} allows us to conclude that $P(\ysz,\yx)$ is a
Gaussian distribution 
\bea
P(\ysz,\yx) & = & G(\delta;\sigma_{tot}^2) \\
& = & \frac{1}{\sqrt{2\pi \sigma_{tot}^2}} \exp\left( - \frac{1}{2}\frac{\delta^2}{\sigma_{tot}^2} \right) \label{eq:flat_lkhd}
\eea
where 
\bea
\delta & = & \ysz - (a+\alpha \yx) \label{eq:x1} \\
\sigma_{tot}^2 & = & \alpha^2\sx^2 + \ssz^2 + \sszx^2.
\eea 
For a power-law distribution $P(\yx) \propto \exp(-\beta \yx)$, we find
\bea
P(\ysz,\yx) & \propto & G(\delta;\sigma_{tot}^2) \nn \\
 & & \hspace{-0.4in} \times \exp\left[ - \beta\left( y_x f_{sz} + \tilde y_xf_x \right)+ \frac{1}{2} \frac{\beta^2}{\alpha^2} \sigma_{tot}^2 f_x f_{sz}  \right]
\eea
where $\tilde y_x$ is the $y_x$ value expected from $\ysz$ from naively inverting the scaling relation, i.e.
\be
\tilde \yx =\alpha^{-1}(\ysz-a).
\ee
$f_x$ and $f_{sz}$ are the fraction of the total variance due to $\yx$ and $\ysz$ respectively,
\bea
f_x & = & \frac{\alpha^2\sx^2}{\sigma_{tot}^2} \\
f_{sz} & = & \frac{\ssz^2+\sszx^2}{\sigma_{tot}^2}.
\eea
Note that for $\beta=0$ (i.e. a flat prior), we recover equation \ref{eq:flat_lkhd}, as we should.
For our fiducial analysis, we will assume $\beta=1.8$, as appropriate for a logarithmic slope of the
cumulative halo mass function $N(M)\propto M^{\gamma}$ for $\gamma=3$, and a self-similar
slope $M \propto Y_X^{3/5}$.   We have verified that setting $\beta=0.0$ does not impact
our result at any significant level, which suggests our results are not sensitive to the details of the
cluster selection function. This is in good agreement with the more detailed treatment of selection
effects of \citet{planck11_local}, who find that selection effects only impacts the slope of the
best-fit relation, and then only at the $\Delta\alpha=0.01$ level, a systematic that is three times smaller
than the statistical uncertainty.

We assume that all galaxy clusters represent independent realizations of the probability distribution
$P(\ysz,\yx)$, so that the likelihood
of the entire data set is simply the product of the individual likelihoods.
The posterior is
\be
\lkhd(a,\alpha,\sszx|\ysz,\yx) \propto P_0(a,\alpha,\sszx) \prod_{clusters} P(\ysz|\yx)
\ee
where $P_0$ is the prior, and the product is over all clusters.  
Note that while we are using $\sszx$ as a model parameter in order to facilitate
physical interpretation of our results, it is the variance $\sszx^2$ that is more physical, so we adopt a flat
prior on $\sszx^2$, corresponding to
\be
P_0(a,\alpha,\sszx) \propto \sszx.
\ee
%

%%%%%%%%%%%%%%%%%%%%%%%%%%%%%%%%%%%%%%%%%%%%%%%%%%%
%%%%%%%%%%%%%%%%%%%%%%%%%%%%%%%%%%%%%%%%%%%%%%%%%%%
%%%%%%%%%%%%%%%%%%%%%%%%%%%%%%%%%%%%%%%%%%%%%%%%%%%
%%%%%%%%%%%%%%%%%%%%%%%%%%%%%%%%%%%%%%%%%%%%%%%%%%%
%%%%%%%%%%%%%%%%%%%%%%%%%%%%%%%%%%%%%%%%%%%%%%%%%%%
%%%%%%%%%%%%%%%%%%%%%%%%%%%%%%%%%%%%%%%%%%%%%%%%%%%

\subsection{Aperture Correlations}

As detailed
above, we have assumed that both $Y_X$ and $\Ysz$ are estimated within the
same aperture, and we have explicitly assumed that the $\Ysz/Y_X$ ratio is radius independent.
This guarantees that the mean relation is always adequately sampled.  However, any noise
in the estimated aperture $\Rf$ necessarily induces correlated noise in the estimated $\Ysz$
and $Y_X$ values for each clusters, and this covariance should in principle be included in the analysis.
In practice, however, this is difficult to achieve because the $Y_X$ aperture adopted is
that of \citet{planck11_local} rather than the $\Rf$ aperture that we would assign based on
the $Y_X$ measurements of \citet{vikhlininetal09}.  This makes developing a self-consistent
model that accounts for this covariance difficult.
We can demonstrate, however, that we expect this effect to be negligible.  

To see this, we go back to equation \ref{eq:model}.  We now let $\bar \yx$ be the true $\yx$
value at the true radius $\Rf$, while $\bar \yx(R)$ is the true $\yx$ value at radius $R$.
Assuming a power-law scaling
\be
\bar \yx(R) = \bar \yx + \gamma_x \ln \left( \frac{R}{\Rf} \right),
\ee
and letting the observed $\Rf$ radius be a power-law as a function of the observed $Y_X$, one has
\be
\ln\left( \frac{R_{obs}}{\Rf} \right) = \tau (\yx - \bar \yx)
\ee
where $\tau$ is some proportionality constant.
Our modified equation \ref{eq:model} is
\be
\yx = \bar \yx(R)+\delta_x,
\ee
so putting it all together, we arrive at
\be
\yx = \bar \yx + \frac{1}{1-\gamma_x \tau}\delta_x.
\ee
The gas density profile of galaxy clusters is such that $\gamma_x=1.2$,
and for a self-similar scaling, $R\propto M^{1/3} \propto Y_X^{1/5}$, so $\tau=0.2$.
Thus, if we self-consistently estimated the radius $\Rf$ from $Y_X$, we would find
that the variance in $Y_X$ increases by $\approx 25\%$.  Since the variance
in $\Ysz$ is much larger than that in $Y_X$, this is not a source of concern.

What about the impact these fluctuations have on $\Ysz$?  A similar argument
as the one above results in
\be
\ysz = \bar \ysz + \frac{\gamma_{sz}\tau}{1-\gamma_{sz}\tau}\delta_x + \delta_{sz}.
\ee
Thus, the fluctuations in $\ysz$ are now explicitly correlated with those in $\yx$.  Note, however,
that $\delta_x \ll \delta_{sz}$ because the SZ measurements are noisier, and that the $\delta_x$
fluctuations are themselves down-weighted by a factor $\approx 0.3$.  This is a very minor perturbation
to the variance in $\ysz$, and the correlation coefficient between $\yx$ and $\ysz$ induced in this manner
is very small indeed.  To first order in $\gamma_{sz}\tau$ and $\gamma_x\tau$ we find
\be
r \approx \gamma_{sz}\tau \frac{ \sx }{\ssz } \approx 0.06.
\ee
Thus, we expect the covariance induced from a self-consistent estimate of $\Rf$ to be negligible in this case.
This independence is further strengthened by the fact that $\Rf$ is not estimated from the \chandra\ data itself,
so while we cannot adequately model the effects of aperture choice on the variance of our measurements,
we do not expect these to impact our results at any significant level.

%%%%%%%%%%%%%%%%%%%%%%%%%%%%%%%%%%%%%%%%%%%%%%%%%%%
%%%%%%%%%%%%%%%%%%%%%%%%%%%%%%%%%%%%%%%%%%%%%%%%%%%
%%%%%%%%%%%%%%%%%%%%%%%%%%%%%%%%%%%%%%%%%%%%%%%%%%%
%%%%%%%%%%%%%%%%%%%%%%%%%%%%%%%%%%%%%%%%%%%%%%%%%%%
%%%%%%%%%%%%%%%%%%%%%%%%%%%%%%%%%%%%%%%%%%%%%%%%%%%
%%%%%%%%%%%%%%%%%%%%%%%%%%%%%%%%%%%%%%%%%%%%%%%%%%%

\section{Results}
\label{sec:results}

We evaluate the posterior on a grid in the 3-dimensional space spanned by the parameters
$a$, $\alpha$, and $\sszx$, where the range of parameters is chosen to as to ensure that it does
not impact our results.  We use $a\in [-0.35,-0.05]$, $\alpha\in[0.8,1.1]$, and 
$\sszx\in[0,0.25]$.\footnote{In section \ref{sec:discussion}, we repeat our analysis using data 
published in \citet{planck11_local}.
When doing so, we always make sure the extent of our multidimensional grid
completely covers the high-likelihood regions in parameters space.}

%%%%%%%%%%%%%%%%%%%%%%%%%%%%%%%%%%%%%%%%%%%%%%%%
%%%%%%%%%%%%%%%%%%%%%%%%%%%%%%%%%%%%%%%%%%%%%%%%

\begin{figure} 
\hspace{-0.35in} \includegraphics[width=3.7in]{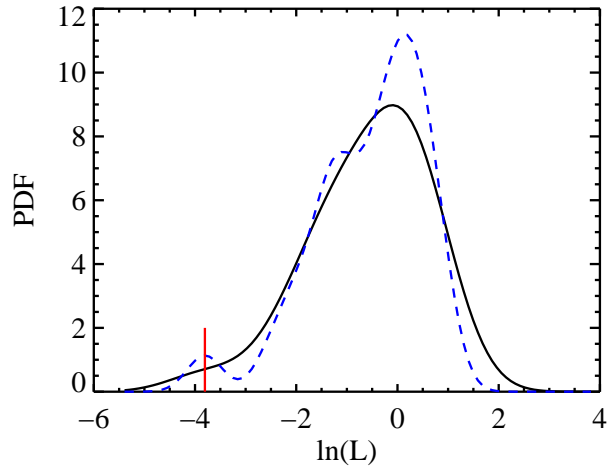}
\caption{The distribution of the log-likelihood value from our cluster sample.  We use our best fit
model to evaluate the log-likelihood of each clusters, and a Gaussian Kernel Density Estimator (KDE)
to estimate the density distribution.  The black line uses 
Silverman's rule of thumb to set the width of the KDE (i.e.  the width of the Gaussian is set to $1.34\sigma/N^{0.2}$ where $N$ is 
the number of samples).  The dashed blue curve halves this width, and
is obviously under-smoothed.  There are no glaring outliers.  The most discrepant cluster is
A2163 (solid red line), which has a low likelihood because of it's extremely large mass.
\\
}
\label{fig:lnlkhd}
\end{figure}

%%%%%%%%%%%%%%%%%%%%%%%%%%%%%%%%%%%%%%%%%%%%%%%%
%%%%%%%%%%%%%%%%%%%%%%%%%%%%%%%%%%%%%%%%%%%%%%%%

%%%%%%%%%%%%%%%%%%%%%%%%%%%%%%%%%%%%%%%%%%%%%%%%
%%%%%%%%%%%%%%%%%%%%%%%%%%%%%%%%%%%%%%%%%%%%%%%%

\begin{figure*} \begin{center}
\includegraphics[width=7in,height=6.1in]{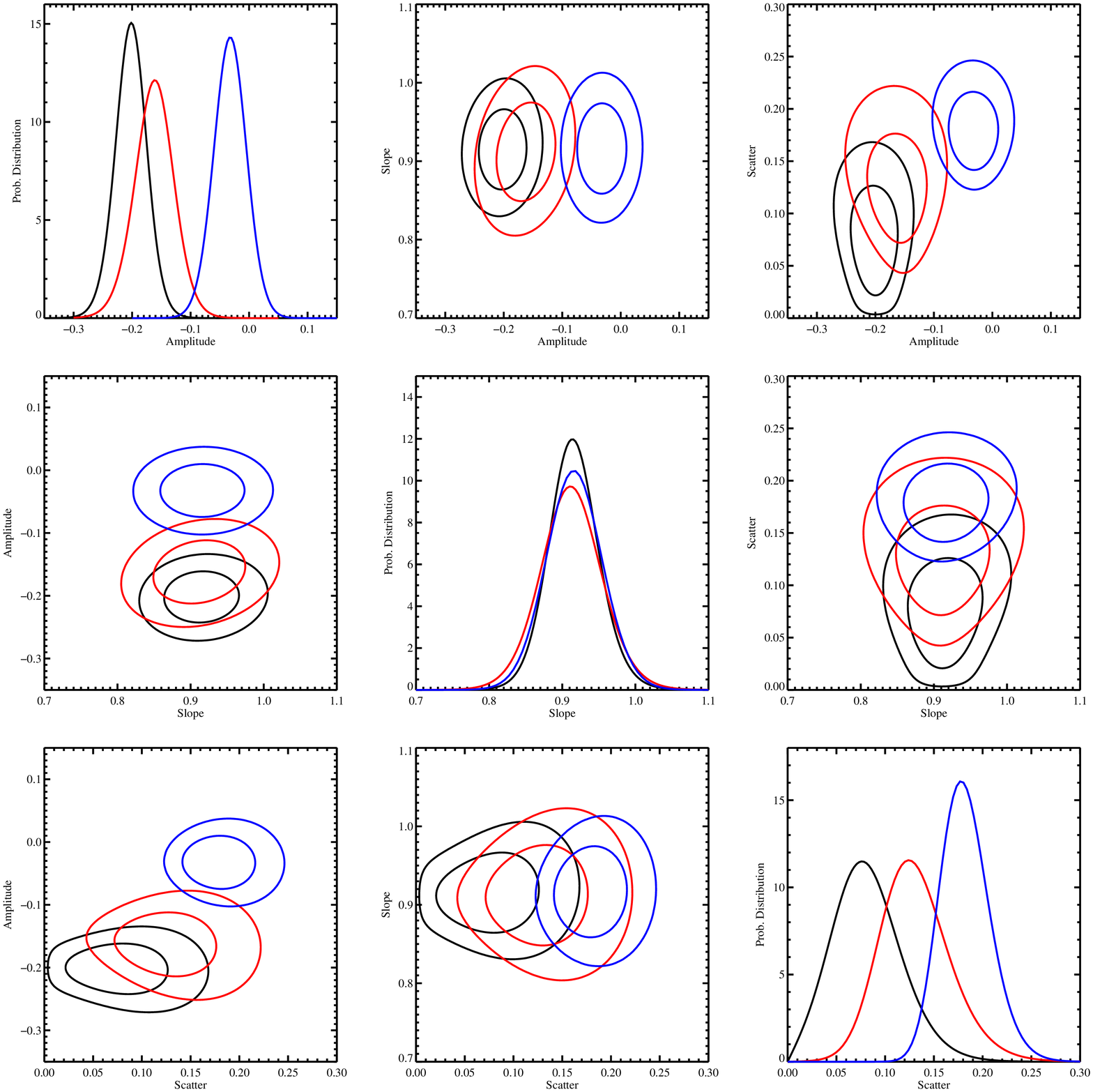}
\caption{Best-fit parameters for the $\Ysz$--$Y_X$ scaling relation.  Black curves are obtained
using \chandra\ data and the cluster sample in Table \ref{tab:clusters}.  Red and blue
curves are using the \citet{planck11_local} data.  The red curves show the results when we restrict our
analysis of the \citet{planck11_local} data to the same galaxy clusters that constitute our \chandra\ sample.
The blue curves include all galaxy clusters in \citet{planck11_local}.
All confidence regions are 68\% and 95\% likelihood contours respectively.
}
\label{fig:fits}
\end{center} 
\end{figure*}

%%%%%%%%%%%%%%%%%%%%%%%%%%%%%%%%%%%%%%%%%%%%%%%%
%%%%%%%%%%%%%%%%%%%%%%%%%%%%%%%%%%%%%%%%%%%%%%%%

Figure \ref{fig:fits} shows our best fit scaling relation.  The best fit amplitude (i.e. the $\Ysz/Y_X$ ratio),
slope, and scatter are
\bea
a & = &-0.202\pm 0.028 \label{eq:yszyx_amp} \\
\alpha & = & 0.916\pm 0.035 \\
\sszx & = & 0.082\pm 0.035.
\eea
In all cases, the quoted values are the mean of the distribution, and 
the error is the standard deviation of the marginalized distribution.  The slope is marginally consistent
with $\alpha=1$ ($\approx 2.3\sigma$ offset).

We have tested the sensitivity of our result to outliers by performing jackknife resamples of our galaxy clusters,
and repeating our analysis.  The resulting mean and standard deviation are
\bea
a & = &-0.202\pm 0.024 \\
\alpha & = & 0.916\pm 0.032 \\
\sszx & = & 0.083\pm 0.021
\eea
The central values and error estimates are essentially identical to those derived from our full likelihood analysis,
demonstrating that there are no large outliers.  This can be furthered evidence by evaluating the
log-likelihood of each individual cluster.  The distribution of log-likelihoods is shown in
Figure~\ref{fig:lnlkhd}.  The least likely cluster is A2163, reflecting the extremely large mass of this system.
By the same token, A2163 significantly increases
the lever-arm with which the slope of the $\Ysz$--$Y_X$ relation can be constrained, and has an important impact
on the fit.
Removing A2163 from the fit results in a slope $\alpha=0.906\pm0.046$.  The error increases
significantly, but the deviation of $\alpha$ from unity is still $2\sigma$. 

\begin{figure} 
\includegraphics[width=3.5in]{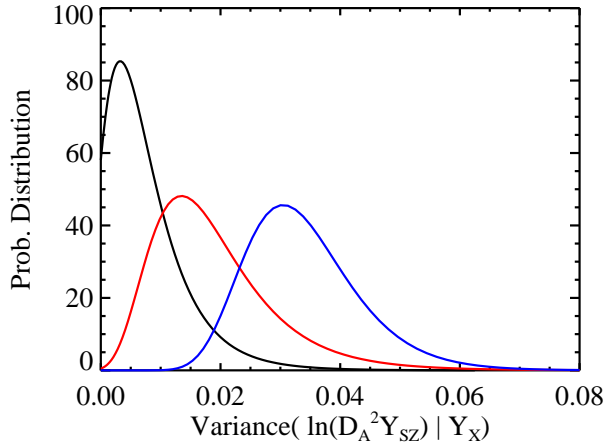}
\caption{Likelihood distribution for the intrinsic variance in $\ln \Ysz$ at fixed $Y_X$.  Black, red, and
blue curves are for the \chandra--\planck\ data, the \xmm--\planck\ data restricted to the subset of
clusters with \chandra\ data, and the full \xmm--\planck\ data set.  
}
\label{fig:var}
\end{figure}

We test whether the scatter of the $\Ysz$--$Y_X$ relation is consistent with zero or not
by focusing not on the scatter, but on the variance, $\sszx^2$, which is the more physical parameter.
The marginalized likelihood for the variance $\sszx^2$ is shown in Figure \ref{fig:var}.  We see
that the data is consistent with zero variance.  We can place an upper limit on the variance,
$\sszx^2 \leq 0.021$ (95\% CL), corresponding to $\sszx \leq 0.144$.

We also explicitly consider the $\Ysz$--$Y_X$ relation with a fixed slope of unity ($\alpha=1$).
The scatter is somewhat increased to $\sszx=0.120\pm0.033$.  The uncertainty in the amplitude also increases, 
reflecting the additional intrinsic  scatter required to fit all the data. We find $a=-0.201\pm 0.032$.
In this context, we also investigate whether the resulting amplitude is sensitive to the dynamical state of the
galaxy clusters.  We split the clusters depending on whether they are relaxed or unrelaxed, using the same
criteria employed in \citet{vikhlininetal09}.  The best fit amplitude for the relaxed and unrelaxed cases
is $a=-0.17\pm 0.04$ and $a=-0.26\pm 0.06$, consistent with no systematic offset between the two.

%%%%%%%%%%%%%%%%%%%%%%%%%%%%%%%%%%%%%%%%%%%%%%%%
%%%%%%%%%%%%%%%%%%%%%%%%%%%%%%%%%%%%%%%%%%%%%%%%

%%%%%%%%%%%%%%%%%%%%%%%%%%%%%%%%%%%%%%%%%%%%%%%%%%%
%%%%%%%%%%%%%%%%%%%%%%%%%%%%%%%%%%%%%%%%%%%%%%%%%%%

%%%%%%%%%%%%%%%%%%%%%%%%%%%%%%%%%%%%%%%%%%%%%%%%%%%
%%%%%%%%%%%%%%%%%%%%%%%%%%%%%%%%%%%%%%%%%%%%%%%%%%%
%%%%%%%%%%%%%%%%%%%%%%%%%%%%%%%%%%%%%%%%%%%%%%%%%%%
%%%%%%%%%%%%%%%%%%%%%%%%%%%%%%%%%%%%%%%%%%%%%%%%%%%
%%%%%%%%%%%%%%%%%%%%%%%%%%%%%%%%%%%%%%%%%%%%%%%%%%%
%%%%%%%%%%%%%%%%%%%%%%%%%%%%%%%%%%%%%%%%%%%%%%%%%%%

\section{Discussion}
\label{sec:discussion}

\subsection{Comparison to Previous Work}

We compare our best fit $\Ysz$--$Y_X$ scaling relation to other empirical investigations of the $\Ysz/Y_X$ ratio.
There are several analysis of the SZ and X-ray properties of galaxy clusters, but most cannot be straight forwardly
compare to ours.
\citet{bonamenteetal08} do not explicitly consider the $\Ysz$--$Y_X$ relation, but simply note that 
an isothermal model with no gas clumping gives a good fit to their data.  This model corresponds
to $\ln(\Ysz/Y_X)=0$, which is in strong conflict with our results. 
We note, however, that they
used significantly smaller apertures ($R_{2500}$), and that their SZ and X-ray estimators are explicitly
linked in the analysis: the SZ data helps constrain the X-ray data model and vice-versa, and the isothermal
$\beta$ model effectively acts as a prior relating the two data sets.
Thus, it is difficult to fairly compare their analysis to ours.

On a similar vein, \citet{comisetal11} used the ACCEPT cluster catalog \citep{cavagnoloetal09} to probe
the $\Ysz$--$Y_X$ relation, but within an overdensity $\Delta=2500$.   In addition, \citet{comisetal11}
do not exclude the core when estimating $T_X$, which is expected to introduce scatter, and to systematically
bias the $Y_X$ measurements relative to what would be measured using core-excluded temperatures.  
Indeed, \citet{comisetal11} find both higher scatter and a higher  $\Ysz/Y_X$ ratio, as we would expect.
A more quantitative comparison with our results, however, is difficult.

The \citet{planck11_local} analysis is the one that is closest in
spirit to ours.  Further, our analysis relies on their SZ-data.
Switching the best fit amplitude reported in that work to the pivot
point adopted here, their best fit scaling relation corresponds to
$a=-0.037 \pm 0.023$ and $\alpha=0.96\pm 0.04$.
 This amplitude is significantly higher than ours at more
than $4.5\sigma$. Their slope is steeper, but consistent with ours.
There is also tension between the recovered scatter of their analysis
and ours, with the $\sszx=0.228$ value reported by
\citet{planck11_local} being ruled out by our data at very high
significance.\footnote{The number quoted in \citet{planck11_local} is
  $0.10\pm 0.010$, but they quote scatter in dex, while we quote the
  scatter in $\ln \Ysz$.}

What drives the difference between our results and those of
\citet{planck11_local}?  Two obvious possibilities that we can readily
explore are the fitting method used to recover the scaling relation,
and the fact that we use a subsample (28 out of 62) of the galaxy
clusters employed in the \citet{planck11_local} analysis.

We test the importance of the fitting method by refitting the
\citet{planck11_local} data with our likelihood method.  We find
$a=-0.032 \pm 0.028$, $\alpha=0.917\pm0.039$, and $\sszx=0.182 \pm
0.025$.  For this analysis, we have used the
data exactly as reported in \citet{planck11_local}, that is, we use
their redshifts, and we assume $h=0.70$, though we note that the $h$
value impact the amplitude $a$ only at the $1\%$ level. The only
exception is cluster A2034, for which the redshift in
\citet{planck11_local} is clearly incorrect \citep[][quote $z=0.151$,
while we use the cD galaxy redshift, $z=0.113$, a very significant
difference]{planck11_local}.  These results are illustrated with the
blue lines in Figure \ref{fig:fits}.  Our best fit amplitude is in
agreement with that quoted in \citet{planck11_local}, though our best
fit value for the intrinsic scatter is lower, with $\Delta \sszx
\approx 0.04$ (partly because we have correct the redshift of cluster A2034),
and our best fit slopes are flatter.
Despite the differences noted above, we can conclude that the fitting method is not responsible
for the difference between our results and those of \citet{planck11_local}.

We test the importance of using a subsample of \citet{planck11_local}
data by analyzing the \citet{planck11_local} data, but restricting our analysis
to the clusters employed in our \chandra\ analysis.
In this case, we find $a=-0.166 \pm
0.035$, $\alpha=0.912 \pm 0.043$, and $\sszx = 0.132 \pm 0.036$.  
We see that the sample selection plays a
significant impact on the results of the recovered $\Ysz/Y_X$ ratio
for the \xmm\ data. 
The higher normalization of the $Y_{X}-\Ysz$ relation from the full \planck\ sample must therefore be driven by the remaining
34 clusters not in the \chandra\ subsample (most of which are at $z>0.1$).
Note that while we have not attempted to replicate the Monte Carlo analysis of \cite{planck11_local}
to try to estimate selection effects, their own tests suggest that there are negligible, having no impact on the amplitude of
the $\Ysz$--$Y_X$ relation, and impacting the slope only at the $\Delta\alpha=0.01$ level, much less than the statistical uncertainty
in the measurement.

Even when restricting ourselves to the same galaxy clusters, however,
the \chandra\ data results in a lower $\Ysz/Y_X$ ratio.  While this
difference is $<1\sigma$ in terms of the statistical error bar, we
note the statistical significance is significantly higher because both
data sets share the same $\Ysz$ data, and it is the $\Ysz$ error that
dominates the error budget.  More specifically, the fact that both
clusters use the exact same $\Ysz$ data implies that the the
difference must be driven entirely by systematic differences in $Y_X$
between the two data sets.

Aside from the differences in amplitude, it is interesting to note the
difference in the intrinsic scatter estimate between the two data
sets.   The simplest interpretation of this result is that
there are unidentified sources of stochasticity in the
\citet{planck11_local} X-ray analysis that result in an overestimate
of the intrinsic scatter of the $\Ysz$--$Y_X$ relation of galaxy
clusters.

Finally, it is worth noting that all of our fits consistently return a
slope $\alpha\approx 0.91$, compared to the $\alpha=0.96$ slope
reported in \citet{planck11_local}.  This difference appears to be due
to the choice of fitting method: fitting our data using the same BCES
method employed by \citet{planck11_local}, we recover $\alpha=0.96$.
Since we find no evidence for non-gaussian scatter intrinsic scatter
in the $\Ysz-Y_{X}$ relation, and furthermore, the observed scatter is
dominated by measurement uncertainties, we believe that using the
maximum likelihood fit is better justified.

The final data set that we compare against is that of
\citet{anderssonetal10}, who find $\ln \Ysz/Y_X = -0.20 \pm 0.1$ using
\chandra\ X-ray data and SPT SZ data with a median cluster redshift $z\approx 0.7$.
This value is essentially in perfect agreement with ours, and is also consistent
with that of \citet{planck11_local}.
The uncertainty in their measurement is significantly larger than the
one reported here or in \citet{planck11_local} both because of a
smaller cluster sample, and because the X-ray data is also shallower
in terms of number of X-ray photons. We conclude that from the
comparison of \planck+\emph{Chandra} and SPT+\emph{Chandra} samples,
there is no evidence for evolution in the $\Ysz-Y_{X}$ relation. It
will be interesting to make the same comparison in the future as both
projects acquire more galaxy clusters with high quality SZ and X-ray
data.

%-------------------------------------------------------------------------
%-------------------------------------------------------------------------

\begin{figure}[t]
\begin{center}
\hspace{-0.2in}\scalebox{1.2}{\plotone{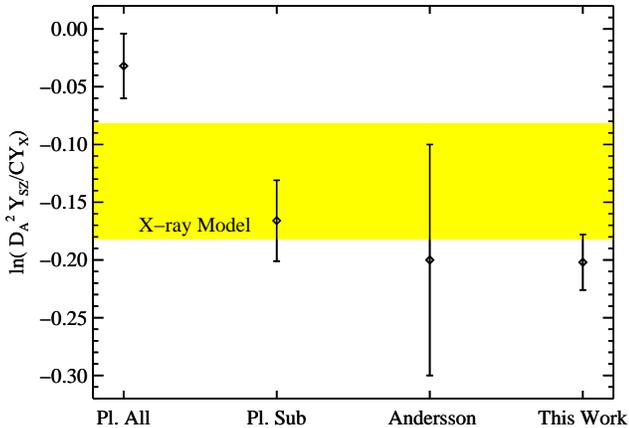}}
\caption{The $\Ysz/Y_X$ ratio, as observed using a variety of data.
The horizontal band is our X-ray predictions model from section \ref{sec:predictions},
the width of which ($\pm 0.05$) reflect our ignorance of clumping corrections.
The ``Pl. All'' data point shows our re-analysis of the \citet{planck11_local} data.
``Pl. Sub'' uses this same data, but restricts the cluster sample to that employed in our
\chandra\ analysis.  The Andersson point 
comes from \citet{anderssonetal10}. 
}
\label{fig:ysz_yx_ratio}
\end{center}
\end{figure}

%-------------------------------------------------------------------------
%-------------------------------------------------------------------------

For completeness, we have also estimated the amplitude of the $\Ysz$--$Y_X$ relation setting $\alpha=1$
for both our data, and the various sub-samples of the \citet{planck11_local} data we considered here.
The likelihood for each of these fits is shown in Figure \ref{fig:lkhd2d}.
In addition, we also computed the best fit amplitude for the low redshift
($z \leq 0.12$) and intermediate redshift ($0.12< z < 0.3$) sub-samples from \citet{planck11_local}.
Our results are summarized in Figure \ref{fig:ysz_yx_ratio}.

%%%%%%%%%%%%%%%%%%%%%%%%%%%%%%%%%%%%%%%%%%%%%%%%%%%
%%%%%%%%%%%%%%%%%%%%%%%%%%%%%%%%%%%%%%%%%%%%%%%%%%%
%%%%%%%%%%%%%%%%%%%%%%%%%%%%%%%%%%%%%%%%%%%%%%%%%%%
%%%%%%%%%%%%%%%%%%%%%%%%%%%%%%%%%%%%%%%%%%%%%%%%%%%
%%%%%%%%%%%%%%%%%%%%%%%%%%%%%%%%%%%%%%%%%%%%%%%%%%%
%%%%%%%%%%%%%%%%%%%%%%%%%%%%%%%%%%%%%%%%%%%%%%%%%%%

\subsection{Comparison to X-ray Expectations}

Our empirical calibration of the $\Ysz-Y_{X}$ relation for $z<0.1$
Planck clusters is somewhat lower than the expectation from the
observed cluster temperature profiles for both \emph{Chandra} and
\emph{XMM} measurements, $\Delta\ln(\Ysz/Y_{X})_{\it XMM} =
-0.09\pm0.035$ and $\Delta\ln(\Ysz/Y_{X})_{\it Chandra} =
-0.127\pm0.024$. Some of this offset can be explained by small-level
nonuniformities of the intracluster gas (gas clumping factor $Q$ in
eq.~\ref{eq:Tratio2}). As per our earlier discussion, 
we set $Q=1.05\pm0.05$ \citep{nagaietal07a}, which brings the
predicted $\Ysz/Y_X$ ratio within a few percent of the observed
value, $\Delta\ln(\Ysz/Y_{X})_{\it XMM} = 0.09$ for the full \xmm--\planck\
cluster sample,
$\Delta\ln(\Ysz/Y_X)_{\it XMM} = -0.042\pm0.035$ for
the \chandra\ sub-sample of galaxy clusters,
and $\Delta\ln(\Ysz/Y_{X})_{\it Chandra} = -0.078\pm0.024$
as estimated from \chandra.
Note that the estimated systematic uncertainty in the clumping correction
is $\pm 0.05$, so all values are in reasonable agreement with expectations,
even though some are in strong statistical tension with one another.

Our results fit within the general picture that has developed 
in the recent past.  
A few years back, a variety of works argued that the SZ
  signal of galaxy clusters was inconsistent with predictions from
  X-ray data \citep{lieuetal06,bielbyshanks07}.  These authors fit
  isothermal $\beta$ models to \rosat\ data in order to predict the
  SZ-signal of galaxy clusters as observed by WMAP, finding that the
  observed SZ signal was significantly lower than that their
  predictions.  There is now ample evidence that $\beta$ models are
  inadequate fits to the intra-cluster gas of galaxy clusters
  \citep[e.g][]{vikhlininetal06,crostonetal08}, and that clusters are
  not isothermal
  \citep[e.g.][]{vikhlininetal06,prattetal07,sunetal09,morettietal11}.
  \citet{hallmanetal07} and \citet{mroczkowskietal09} explicitly
  demonstrated that isothermal model fits to X-ray data leads one to
  over-predict the SZ signal of galaxy clusters.  Using more realistic
  density and temperature profiles, one finds good agreement between
  SZ and X-ray observables
  \citep{atrio-barandelaetal08,komatsuetal08,mroczkowskietal09,diegopartridge10,zwartetal11,
  melinetal11,planck11_local,planck11_xray}.  The results summarize above fit well within this new
  wave of works.

%%%%%%%%%%%%%%%%%%%%%%%%%%%%%%%%%%%%%%%%%%%%%%%%%%%
%%%%%%%%%%%%%%%%%%%%%%%%%%%%%%%%%%%%%%%%%%%%%%%%%%%
%%%%%%%%%%%%%%%%%%%%%%%%%%%%%%%%%%%%%%%%%%%%%%%%%%%
%%%%%%%%%%%%%%%%%%%%%%%%%%%%%%%%%%%%%%%%%%%%%%%%%%%
%%%%%%%%%%%%%%%%%%%%%%%%%%%%%%%%%%%%%%%%%%%%%%%%%%%
%%%%%%%%%%%%%%%%%%%%%%%%%%%%%%%%%%%%%%%%%%%%%%%%%%%

\section{Summary and Conclusions}

Using SZ data from \planck\ \citep{planck11_local} and X-ray data from
\chandra\ \citep{vikhlininetal09}, we have estimated the $\Ysz$--$Y_X$
scaling relation.  Assuming a power-law of the form in equation
\ref{eq:powerlaw}, we find $\ln(\Ysz/Y_X) = a = -0.202\pm 0.024$,
$\alpha=0.916\pm 0.032$, and $\sszx=0.083\pm0.021$.  The scatter
constraint is prior-driven, and the variance in the relation is
consistent with zero intrinsic scatter.  The slope $\alpha$ differs
from unity at the $\approx 2.3\sigma$ level, which is interesting but
not yet significant.  We find no significant difference in the scaling
relations when comparing relaxed and unrelaxed systems.

Our mean $\Ysz/Y_X$ is significantly lower than that reported from the
\emph{XMM} and \planck\ analysis of the full sample of 62 clusters ($a=-0.032\pm0.028$).
However, when we repeat the \planck\ and \xmm\ analysis restricting
ourselves to the same 28 systems that compose the \chandra\ subsample,
we find results that are in significantly better agreement ($a=-0.166\pm 0.036$),
demonstrating the original tension is driven by the remaining 34 (mostly $z\geq 0.1$)
clusters in the full \planck\ and \xmm\ cluster sample.

Compared with the \emph{XMM-Newton} analysis, we find a significantly
lower scatter in the $\Ysz-Y_{X}$ relation, $\sigma_{sz|x}=0.083\pm 0.021$ for the \chandra--\planck\
data, versus $\sigma_{sz|x}= 0.182\pm 0.025$ for the full \xmm--\planck\ sample.  The scatter for the
full \xmm--\planck\ sample is at least partly driven by the systematic differences between the \chandra\ sub-sample
of galaxy clusters, and the remaining systems.  The \xmm--\planck\ scatter for the \chandra\ subsample is
lower $\sigma_{sz|x}=0.132\pm 0.036$ and consistent with the scatter estimated from \chandra\ data.
Finally, we note the \chandra\ data is fully consistent with no intrinsic scatter in the $\Ysz$--$Y_X$
relation: our central values are prior-driven.

Both the \emph{Chandra} and \emph{XMM} values of $\Ysz/Y_{X}$ for
$z<0.1$ are significantly lower than $a=-0.075$ expected from the
observed cluster temperature profiles along (\S~\ref{sec:predictions}). We
believe that a significant fraction of that difference can be
attributed to some level of gas clumping, with $Q\simeq1.05$ including substructure masking \citep{nagaietal07a}.
Any remaning difference is
within the range of current absolute accuracy of the X-ray temperature
measurements and systematic uncertainty in the clumping factor correction.

In summary, we find that the observed $Y_{X}-\Ysz$ relation from the
joint \planck\ and \emph{Chandra} analysis is in good agreement with the
theoretical expectations --- it shows very low intrinsic scatter, a
slope close to unity, and the amplitude is consistent with physically plausible
levels of gas clumping.

In conjunction with the work in \citet{vikhlininetal09}, our results can be used to self-consistently estimate 
the multi-variate scaling relation of galaxy clusters.  In an upcoming work, 
we use these results to study the $L_X$--$M$ and $\Ysz$--$M$ scaling relations of galaxy clusters as 
determined from X-ray data, and compare these to those derived from the optically selected
maxBCG galaxy cluster catalog \citep{koesteretal07a}.

\acknowledgements The authors would like to thank August Evrard, James Bartlett, and Eli Rykoff
for useful discussions, and comments on draft versions of the manuscript.  We also thank Monique Arnaud
and Gabriel Pratt for useful criticism on an earlier draft of this work.
ER is funded by NASA through the Einstein Fellowship Program, grant PF9-00068.
AV is funded by NASA grant GO1-12168X and contract NAS8-39073.
SM is supported by the KICP through the NSF grant PHY-0551142 and an endowment from the Kavli Foundation.

\bibliographystyle{apj}
\bibliography{mybib}

\newcommand\AAA[3]{{A\& A} {\bf #1}, #2 (#3)}
\newcommand\PhysRep[3]{{Physics Reports} {\bf #1}, #2 (#3)}
\newcommand\ApJ[3]{ {ApJ} {\bf #1}, #2 (#3) }
\newcommand\PhysRevD[3]{ {Phys. Rev. D} {\bf #1}, #2 (#3) }
\newcommand\PhysRevLet[3]{ {Physics Review Letters} {\bf #1}, #2 (#3) }
\newcommand\MNRAS[3]{{MNRAS} {\bf #1}, #2 (#3)}
\newcommand\PhysLet[3]{{Physics Letters} {\bf B#1}, #2 (#3)}
\newcommand\AJ[3]{ {AJ} {\bf #1}, #2 (#3) }
\newcommand\aph{astro-ph/}
\newcommand\AREVAA[3]{{Ann. Rev. A.\& A.} {\bf #1}, #2 (#3)}

\appendix

\section{Cluster Data}

We summarize the data employed in our analysis in table \ref{tab:clusters}.

%%%%%%%%%%%%%%%%%%%%%%%%%%%%%%%%%%%%%%%%%%%%%%%%
%%%%%%%%%%%%%%%%%%%%%%%%%%%%%%%%%%%%%%%%%%%%%%%%

\begin{figure*} \begin{center}
\includegraphics[width=7in,height=6.1in]{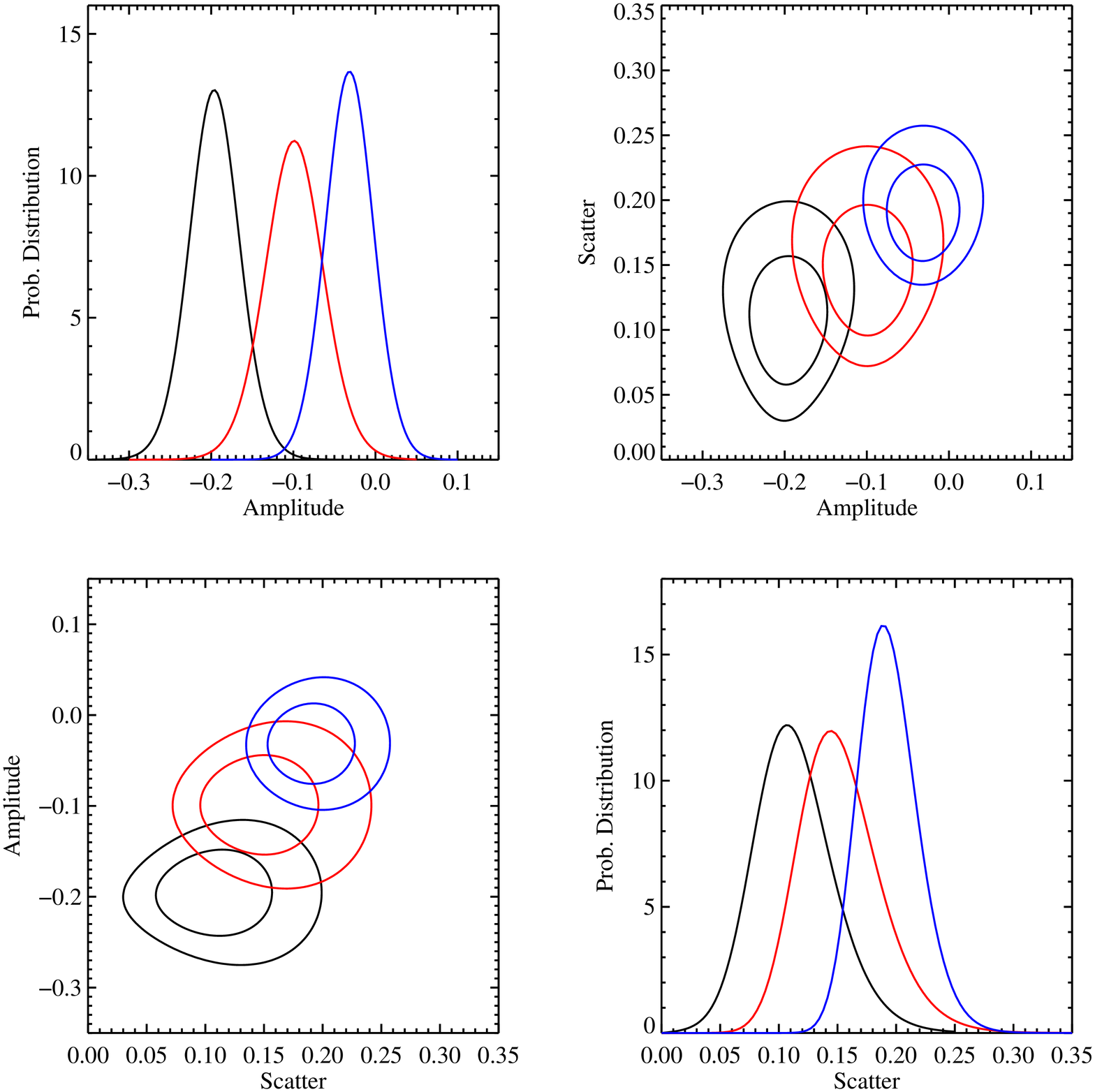}
\caption{Best-fit parameters for the $\Ysz$--$Y_X$ scaling relation assuming a slope of unity.  
Black curves are obtained
using \chandra\ data and the cluster sample in Table \ref{tab:clusters}.  Red and blue
curves are using the \citet{planck11_local} data.  The red curves show the results when we restrict our
analysis of the \citet{planck11_local} data to the same galaxy clusters that constitute our \chandra\ sample.
The blue curves include all galaxy clusters in \citet{planck11_local}.
All confidence regions are 68\% and 95\% likelihood contours respectively.
}
\label{fig:lkhd2d}
\end{center} 
\end{figure*}

%%%%%%%%%%%%%%%%%%%%%%%%%%%%%%%%%%%%%%%%%%%%%%%%
%%%%%%%%%%%%%%%%%%%%%%%%%%%%%%%%%%%%%%%%%%%%%%%%

%%%%%%%%%%%%%%%%%%%%%%%%%%%%%%%%%%%%%%%%
%%%%%%%%%%%%%%%%%%%%%%%%%%%%%%%%%%%%%%%%

\begin{deluxetable}{lcccccc}
\tablewidth{0pt}
\tablecaption{Cluster Data}
\tablehead{Name & P11 Redshift & Redshift & $D_A^2\Ysz$ & $CY_X$ (\chandra) & $C Y_X$ (\xmm) & Merger?}
\startdata
A85 & 0.052 &0.0557 & $5.05 \pm 0.54$ & $6.19 \pm 0.16$ & $5.37 \pm 0.22$ & ---  \\
A119 & 0.044 & 0.0445 & $2.60 \pm 0.29$ & $3.55 \pm 0.13$ & $3.42 \pm 0.16$ & \checkmark \\
A401 & 0.075 & 0.0743 & $7.71 \pm 0.74$ & $10.86 \pm 0.47$ & $10.42 \pm 0.75$ & --- \\
A3112 & 0.070 & 0.0759 & $1.98 \pm 0.33$ & $3.38 \pm 0.21$ & $2.82 \pm 0.11$ & --- \\
A3158 & 0.060 & 0.0583 & $3.14 \pm 0.27$ & $3.10 \pm 0.07$ & $3.73 \pm 0.15$ & --- \\
A478 & 0.088 & 0.0881 & $8.71 \pm 0.76$ & $10.17 \pm 0.58$ & $9.59 \pm 0.39$ & --- \\
A3266 & 0.059 & 0.0602 & $8.84 \pm 0.69$ & $11.89 \pm 0.29$ & $10.07 \pm 0.36$ &\checkmark \\
A3376 & 0.045 & 0.0455 & $0.97 \pm 0.19$ & $1.64 \pm 0.24$ & $1.34 \pm 0.06$ & \checkmark \\
A1413 & 0.143 & 0.1429 & $6.99 \pm 0.76$ & $7.83 \pm 0.48$ & $7.58 \pm 0.12$ & ---  \\
ZwCl1215 & 0.077 & 0.0767 & $4.32 \pm 0.66$ & $5.54 \pm 0.21$ & $5.72 \pm 0.30$ & ---\\
A3528s & 0.053 & 0.0574 & $2.41 \pm 0.33$ & $2.00 \pm 0.15$ & $1.62 \pm 0.10$ & --- \\
A1644 & 0.047 & 0.0475 & $2.41 \pm 0.39$ & $3.08 \pm 0.30$ & $2.80 \pm 0.13$ & \checkmark \\
A1650 & 0.084 & 0.0823 & $4.00 \pm 0.55$ & $3.62 \pm 0.15$ & $3.67 \pm 0.08$ & ---\\
A1651 & 0.084 & 0.0853 & $3.50 \pm 0.58$ & $5.32 \pm 0.29$ & $4.12 \pm 0.12$ & ---\\
A1689 & 0.183 & 0.1828 & $12.93 \pm 1.42$ & $13.42 \pm 0.71$ & $12.41 \pm 0.22$ & --- \\
A3558 & 0.047 & 0.0469 & $3.95 \pm 0.47$ & $4.04 \pm 0.36$ & $4.50 \pm 0.18$ & \checkmark \\
A1795 & 0.062 & 0.0622 & $4.37 \pm 0.38$ & $5.34 \pm 0.18$ & $6.78 \pm 0.28$ & --- \\
A1914 & 0.171 & 0.1712 & $9.47 \pm 0.85$ & $13.29 \pm 0.68$ & $12.43 \pm 0.31$ & --- \\
A2034 & 0.151 & 0.1130 & $4.27 \pm 0.58$ & $5.11 \pm 0.15$ & $6.01 \pm 0.14$ & --- \\
A2029 & 0.078 & 0.0779 & $7.65 \pm 0.66$ & $11.48 \pm 0.53$ & $12.13 \pm 0.65$ & --- \\
A2065 & 0.072 & 0.0723 & $3.72 \pm 0.48$ & $4.35 \pm 0.16$ & $4.52 \pm 0.23$ & \checkmark \\
A2163 & 0.203 & 0.2030 & $43.01 \pm 1.99$ & $60.93 \pm 3.21$ & $59.75 \pm 2.14$ & \checkmark  \\
A2204 & 0.152 & 0.1511 & $10.39 \pm 0.94$ & $12.96 \pm 1.00$ & $11.88 \pm 0.39$ & --- \\
A2256 & 0.058 & 0.0581 & $6.73 \pm 0.38$ & $8.94 \pm 0.43$ & $7.02 \pm 0.33$ & \checkmark \\
A2255 & 0.081 & 0.0800 & $4.81 \pm 0.37$ & $4.97 \pm 0.14$ & $4.81 \pm 0.15$ & --- \\
RX J1720 & 0.164 & 0.1593 & $5.68 \pm 0.72$ & $6.45 \pm 0.38$ & $5.69 \pm 0.14$ & --- \\
A2390 & 0.231 & 0.2302 & $15.61 \pm 1.22$ & $19.49 \pm 1.56$ & $19.26 \pm 0.58$ & --- \\
A3921 & 0.094 & 0.0941 & $3.41 \pm 0.28$ & $3.75 \pm 0.15$ & $3.10 \pm 0.08$ & --- \\
\enddata
\label{tab:clusters}
\tablecomments{The units for $D_A^2\Ysz$ and $CY_X$ are $10^{-5}\ \Mpc^2$.  The $Y_X$ values from \xmm\ are
drawn from \citet{planck11_local}.  In addition, both the $\Ysz$ and $Y_X$ data of \citet{planck11_local}
are corrected to our fiducial cosmology ($h=0.72$) and to our cluster redshifts.  The P11 redshift column quotes the
redshifts as they appear in \citet{planck11_local}.}
\end{deluxetable}

%%%%%%%%%%%%%%%%%%%%%%%%%%%%%%%%%%%%%%%%
%%%%%%%%%%%%%%%%%%%%%%%%%%%%%%%%%%%%%%%%

\end{document}